\documentclass[%
 reprint,
 amsmath,amssymb,
 aps,mathabx,
]{revtex4-1}

\usepackage{graphicx}
\usepackage{dcolumn}

\usepackage{bm}
\usepackage{xcolor}

\begin{document}

\preprint{AIP/123-QED}

\title{A Strategy to Identify Materials Exhibiting a Large Nonlinear Phononics Response:\\ Tuning the Ultrafast Structural Response of LaAlO$_3$ with Pressure}

\author{Jeffrey Z. Kaaret}
\affiliation{ %
School of Applied and Engineering Physics, \\Cornell University, Ithaca, New York 14853, USA
}%
\author{Guru Khalsa}
\author{Nicole A. Benedek}
\email{nbenedek@cornell.edu}
\affiliation{%
Department of Materials Science and Engineering, Cornell University, Ithaca, New York 14853, USA
}%


\begin{abstract}
We use theory and first-principles calculations to investigate how structural changes induced by ultrafast optical excitation of infrared-active phonons change with hydrostatic pressure in LaAlO$_3$. Our calculations show that the observed structural changes are sensitive to pressure, with the largest changes occurring at pressures near the boundary between the cubic perovskite and rhombohedral phases. We rationalize our findings by defining a figure of merit that depends only on intrinsic materials quantities, and show that the peak response near the phase boundary is dictated by different microscopic materials properties depending on the particular phonon mode being excited.
Our work demonstrates how it is possible to systematically identify materials that may exhibit particularly large changes in structure and properties due to optical excitation of infrared-active phonons.

\end{abstract}

\pacs{Valid PACS appear here}
\maketitle

\section{Introduction} \label{sec:introduction}
The past two decades have witnessed significant developments in our ability to control the functional properties of materials with light. In particular, the availability of bright mid-infrared and THz laser sources has made possible experiments involving the mode-selective optical excitation of one or more vibrational (phonon) modes of a material, a phenomenon explored in early theoretical papers\cite{maradudin_ionic_1970,wallis_ionic_1971,humphreys_ionic_1972,martin_ionic_1974,mills_ionic_1987} but only relatively recently exploited in experiments in a systematic way. A distinguishing feature of such experiments is that they involve optical excitation -- and subsequent symmetry breaking -- of the \emph{lattice}, as opposed to the electrons. This mechanism, now known as nonlinear phononics,\cite{forst11,forst15} has been shown to enable dynamical control of superconductivity,\cite{mankowsky14} magnetism,\cite{khalsa18,radaelli18,stupakiewicz21,afanasiev21} ferroelectricity\cite{subedi15,mankowsky17,nova19} and metal-insulator transitions.\cite{rini07}

This control of material properties via the nonlinear phononics mechanism is mediated by an anharmonic coupling between different phonon modes. Specifically, optical excitation of an infrared-active phonon can induce displacements of Raman-active phonons if the crystallographic symmetry of the material allows the modes to be anharmonically coupled in a particular way. The coupling terms that have perhaps been the most investigated to date have the form $AQ_RQ_{IR}^2$, where $Q_R$ is the amplitude of the Raman-active phonon and $Q_{IR}$ is the amplitude of the infrared-active phonon, or $AQ_RQ_{IR,1}Q_{IR,2}$, where $Q_{IR,1}$ is the amplitude of one infrared-active phonon and $Q_{IR,2}$ is the amplitude of a different infrared-active phonon that may be polarized along a different direction.\cite{juraschek17} The magnitude of the coefficient $A$ indicates how strongly coupled the particular infrared-active and Raman-active modes are, and its value is material and mode-dependent.

The nonlinear phononics mechanism has the potential to both dynamically enhance the functional properties of materials and to induce new properties not present in the equilibrium crystal structure. What particular characteristics might indicate that a given material has the potential to exhibit a strong nonlinear phononics response? The answer to this question is currently unknown. Since the change in material properties is primarily understood to be mainly due to the displacement of Raman modes, it follows that the larger the average Raman displacement, the greater the distortion of the equilibrium crystal structure and, presumably, a more significant change in material properties. Hence, given the form of the anharmonic coupling terms above, it is not unreasonable to assume that a large magnitude of $A$, that is, a strong coupling between the optically excited infrared-active mode and the Raman modes, is a prerequisite for a strong nonlinear phononics response. However, there are a number of other variables that affect the response, such as the force constants of the phonon modes involved and the strength of the coupling between the infrared-active mode and the initial light pulse, for example. What are the key microscopic materials factors that dictate the magnitude of the nonlinear phononics response in a given material, and can the response be controlled or enhanced with pressure or epitaxial strain in a thin-film?

We use theory and first-principles density functional theory calculations to explore the questions posed above in the perovskite LaAlO$_3$ in the rhombohedral phase, a material which has been the focus of several recent nonlinear phononics experiments.\cite{caviglia2020LaAlO3, johnson2021LaAlO3} We derive a figure of merit for the average Raman mode displacement in terms of intrinsic materials quantities, such as the coupling coefficient $A$, the force constants of the infrared-active and Raman modes, and the coupling of the infrared-active mode with light. We demonstrate how the figure of merit varies as a function of hydrostatic pressure and decompose the variation into contributions from the microscopic materials properties that define it. Our key findings are that 1) the figure of merit is small and insensitive to pressure for most IR-Raman mode couplings in LaAlO$_3$, and 2) for those mode couplings with a large and pressure-tunable figure of merit, the maximal values of the figure of merit are determined primarily by either a counterbalance between the coupling coefficient $A$ and force constant of the Raman mode, or simply the force constant of the Raman mode. Our results suggest that the factors that give rise to a large nonlinear phononics response are not just material dependent but also depend on the specific phonon modes being excited \emph{within} a given material.  The analysis presented below is general and likely applies to the broader family of perovskites, which are often the subject of nonlinear phononics experiments.

\section{Deriving the Figure Of Merit}
Let us start by considering the lowest-order lattice energy $U$ for the nonlinear phononics process. Without loss of generality, we simplify the analysis by focusing on centrosymmetric crystals since they have infrared-active and Raman-active modes with distinct symmetries. The lattice energy takes the form: \cite{subedi14,fechner16,juraschek17,khalsa18,radaelli18}

\begin{equation} \label{eqn:Potential_energy}
        U  =  \frac{1}{2} K_{IR} Q_{IR}^2  +  \frac{1}{2} K_{R} Q_{R}^2  -  A Q_{R} Q_{IR}^2  -  \widetilde{Z}^*Q_{IR}\vec{E}(t). 
\end{equation}
Here, $K_{IR}$ ($K_R$) is the force constant of the IR-active (Raman) mode, $Q_{IR}$ ($Q_R$) is the corresponding IR-active (Raman) mode amplitude, and the final term describes the coupling between the polarization change in the crystal (defined as $\Delta\vec{P} = \widetilde{Z}^*Q_{IR}$, where $\widetilde{Z}^*$ is the mode-effective charge of the IR-active mode\cite{gonze1997dynamical}) and the electric field $\vec{E}$ of the incoming light pulse. Certain symmetry conditions must be satisfied in order for $A$ to be non-zero. The whole invariant must transform like $A_{1g}$ (or $\Gamma_1^+$, the identical representation) in order to be an allowed term in the lattice energy and hence $Q_R$ must transform like the symmetric part of the direct product of the irreducible representation (irrep) associated with $Q_{IR}$. If $Q_{IR}$ transforms like a one-dimensional irrep, then $Q_R$ must transform like $A_{1g}$ in order for $A$ to be non-zero. However, if $Q_{IR}$ transforms like a two- or higher-dimensional irrep, then $A$ can be non-zero even if $Q_R$ does not transform like $A_{1g}$. The ISOTROPY Software Suite\cite{ISOTROPYGeneral, ISOTROPYInvariants} or the Bilbao Crystallographic Server\cite{Bilbao2011General1, Bilbao2006General2, Bilbao2006General3} can be very helpful in determining allowed mode symmetries for nonlinear phononics coupling.  In this work we consider the optical excitation of a single IR-active mode only, and its coupling to a single Raman mode in order to simplify the analysis in the following sections (in previous work\cite{khalsa18} this scenario was referred to as the two-mode model). 
However, in general there will be dynamic coupling between the optically excited IR-active mode and all other modes of the same symmetry (there will similarly be dynamic coupling between all Raman modes of the same symmetry). This coupling can have a significant effect on the dynamics if, for example, there is more than one IR-active mode within the envelope of the light pulse. See. Ref. \onlinecite{khalsa18} for further details.

The equation of motion for the Raman mode is obtained by taking a derivative of $U$ with respect to the phonon coordinate $Q_R$ for a fixed electric field direction,

\begin{equation} \label{eqn:R_Forces}
        F_R = \frac{ - \partial U}{\partial Q_R} = - K_R Q_R + A Q_{IR}^2,
\end{equation}

\noindent
where $F_R$ is the force on the Raman mode. We derive an expression for the Raman amplitude by setting the force to zero and rearranging Equation \ref{eqn:R_Forces} to give,
\begin{equation}
Q_{R}  =  \frac{ A Q_{IR}^2 }{K_R }.
\label{peakR}
\end{equation}
We now turn to formulating an expression for the average Raman displacement. The IR mode amplitude will be maximized when its frequency is equal to that of the incoming light pulse (resonant excitation), that is,

\begin{align}
        \frac{1}{2}K_{IR} \overline{Q}_{IR}^2 & = \widetilde{Z}^* \overline{Q}_{IR}\widetilde{E} \nonumber \\
        \overline{Q}_{IR} & = \frac{2 \widetilde{Z}^* \widetilde{E}}{K_{IR}}.
        \label{peakIR1}
    \end{align}

Here, $\overline{Q}_{IR}$ is the peak IR amplitude and $\widetilde{E}$ is the effective electric field experienced by the IR phonon and is given by,
\begin{equation}
    \widetilde{E} = E_0 \tau f_{IR},
    \label{efield}
\end{equation}
where $E_0$ is the peak electric field in MV/cm, $\tau$ is the pulse width in ps\cite{note2} and $f_{IR}$ is the frequency of the IR-active mode. There is also a geometric factor associated with the pulse shape (Gaussian, square etc), the effect of which we ignore here. Note that we assume that the pulse width is shorter than the characteristic damping time and so we have ignored damping in deriving the figure of merit. Dielectric screening will reduce the field inside the crystal by a factor of $\epsilon_{\infty}$, at minimum, and in general high-frequency IR modes would also screen the lower-frequency modes. We neglect this screening here although it could be straightforwardly incorporated into the figure of merit. Finally, since the motion of the IR mode is approximately sinusoidal,\cite{note} we can relate the time average of the squared amplitude to the peak amplitude, that is,
\begin{equation}
    \frac{1}{2}\overline{Q}_{IR}^2 = \langle Q_{IR}^2 \rangle.
    \label{peakIR2}
\end{equation}
The figure of merit for the average Raman displacement is obtained by substituting Equations \ref{peakIR1}, \ref{efield} and \ref{peakIR2} into the time average of Equation \ref{peakR} to obtain,
\begin{align}
    \langle Q_R \rangle & = \frac{A\langle Q_{IR}^2 \rangle}{K_R} \nonumber \\
    & = \frac{2A\widetilde{Z}^{*2}}{K_RK_{IR}^2}E_0^2\tau^2 f^2_{IR}.
    \label{alpha1}
\end{align}
Equation \ref{alpha1} naturally separates into contributions involving only intrinsic material quantities (contained in the fraction), and extrinsic experimental parameters (electric field). Dividing both sides by the extrinsic parameters leaves us with an expression purely in terms of intrinsic materials properties:

\begin{equation}
    \frac{\langle Q_R \rangle}{E_0^2\tau^2 } = \frac{2A\widetilde{Z}^{*2} f^2_{IR}}{K_RK_{IR}^2}.
    \label{alpha2}
\end{equation}

\begin{figure}[h]
    \centering
    \includegraphics[width=8cm]{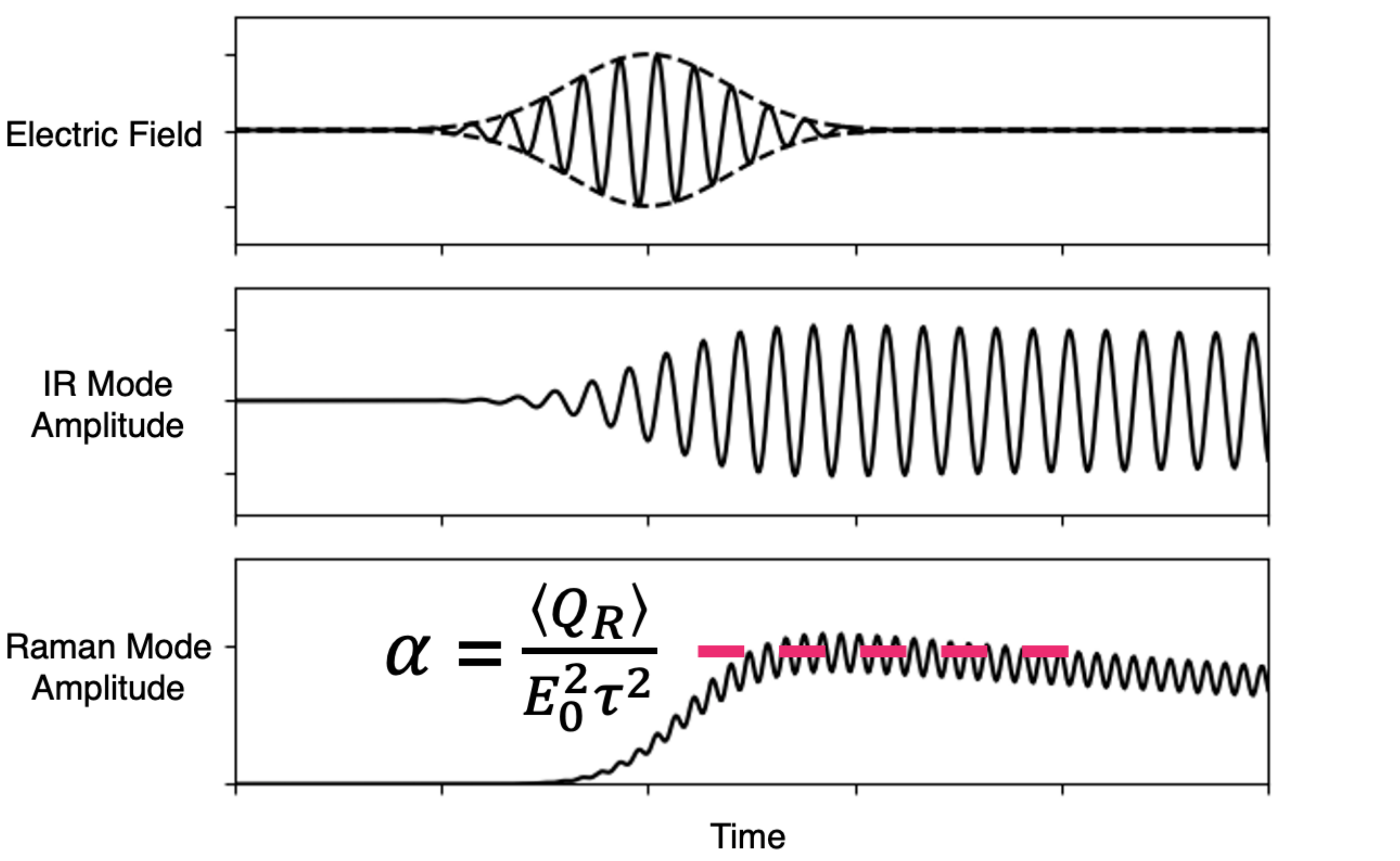}
    \caption{ A schematic illustrating the nonlinear phononics mechanism. The top panel is the incident electric field (with dashed lines indicating a Gaussian envelope), the middle panel shows the resonantly excited IR-active phonon coordinate, and the bottom panel shows the Raman-active mode that is displaced due to anaharmonic coupling with the excited IR mode. The dashed line indicates the average Raman displacement. The high frequency oscillation of the Raman mode is a doubling of the IR mode frequency due to the $AQ_{IR}^2$ force on the Raman mode and is expected from this model. 
}
    \label{fig:FOM_example}
\end{figure}

A similar expression appears in Ref. \onlinecite{forst11}. Equation \ref{alpha2} has units of length per peak electric field squared per pulse width ($\tau$) squared. It is important to note that $\langle Q_R \rangle$ denotes the average Raman \emph{amplitude} for the whole \emph{mode} and not the displacement of any one particular atom. The form of the figure of merit is general and applies to any  crystal with a lattice potential such as that shown in Equation \ref{eqn:Potential_energy}; it is not exclusive to LaAlO$_3$.  We refer to the right-hand side of Equation \ref{alpha2} as $\alpha$, the figure of merit, for the remainder of the manuscript. Figure \ref{fig:FOM_example} shows the derived figure of merit schematically.

Inspection of Equation \ref{alpha2} immediately reveals how $\alpha$, and therefore the average quasistatic Raman displacement, may be maximized in a nonlinear phononics experiment (aside from increasing the peak electric field). The peak Raman displacement is enhanced for small force constants of the Raman  mode ($K_R$) and IR mode ($K_{IR}$). The average Raman displacement will also be maximized if the coupling between the Raman and IR-active modes is strong ($A$ is large) and if the coupling between the IR-active mode and the incoming light pulse is strong (large $\widetilde{Z}^*$).\cite{note3} How does $\alpha$ change with pressure for LaAlO$_3$, and which intrinsic crystal quantities make the dominant contributions to $\alpha$ for specific mode couplings? 

\section{First-Principles Calculations} \label{sec:methods}
The calculations in this work were performed using density functional theory and the PBEsol+U exchange-correlation functional,\cite{Perdew2008PBEsol} as implemented in VASP.5.4.1.\cite{Kresse1996VASP} The electrons included in the valence of the projector augmented wave potentials\cite{Kresse1999VASPPOTS} were: 5$s^2$5$p^6$5$d^1$6$s^2$ for La, 3$s^2$ 3$p^1$ for Al and 2$s^2$2$p^4$ for O. We used a 10-atom rhombohedral cell for all our calculations. Good convergence of the phonon frequencies (within 5 cm$^{-1}$ at standard pressure) was obtained at a plane wave cutoff of 1000 eV and Monkhorst-Pack grid of 6$\times$6$\times$6 k-points compared to higher plane wave cutoffs and denser grids. We found that the frequencies of the softest phonons in LaAlO$_3$ are quite sensitive to the choice of plane wave cutoff, necessitating the use of higher cutoffs than perhaps might otherwise be necessary. Even though the nominal 3+ valence of the lanthanum atom means that the $f$ orbitals are empty, previous work has shown that these states lie too low in energy in standard DFT, potentially leading to spurious mixing with other states. To correct for this, we used a value\cite{JohnsonWilke2013LDADUU} of $U-J$ = 10.32 eV for the on-site Coulomb interaction for the lanthanum $f$ orbitals.\cite{Dudarev1998LDAUU} A force convergence tolerance of 5.0$\times 10^{-4}$ eV/\AA \space was used for all calculations. Our fully relaxed lattice constants are in good agreement with experiment ($a$ = 5.38645 \AA \space and $c$ = 13.1517 \AA \space in the hexagonal setting compared with $a$ = 5.35977  \AA \space and c = 13.0860 \AA \space at 4.2 K from neutron diffraction\cite{Hayward2005Temperature} and $a$ = 5.3638 \AA~ and $c$ = 13.1091 \AA~ at 300 K\cite{lehnert00}). Phonon frequencies, eigenvectors and Born effective charges were calculated using density functional perturbation theory,\cite{Baroni2001DFPT} as implemented in VASP. The mode-effective charge $\widetilde{Z}^*$ was calculated for each IR phonon as defined by Gonze and Lee.\cite{gonze1997dynamical} Anharmonic coupling coefficients were calculated using finite displacements for the modes of interest (we used +15 to -15 picometer displacements of the relevant modes).

\section{Results}
\subsection{Structural Parameters of LaAlO$_3$ Under Pressure}
\begin{figure}
    \includegraphics[width=8cm]{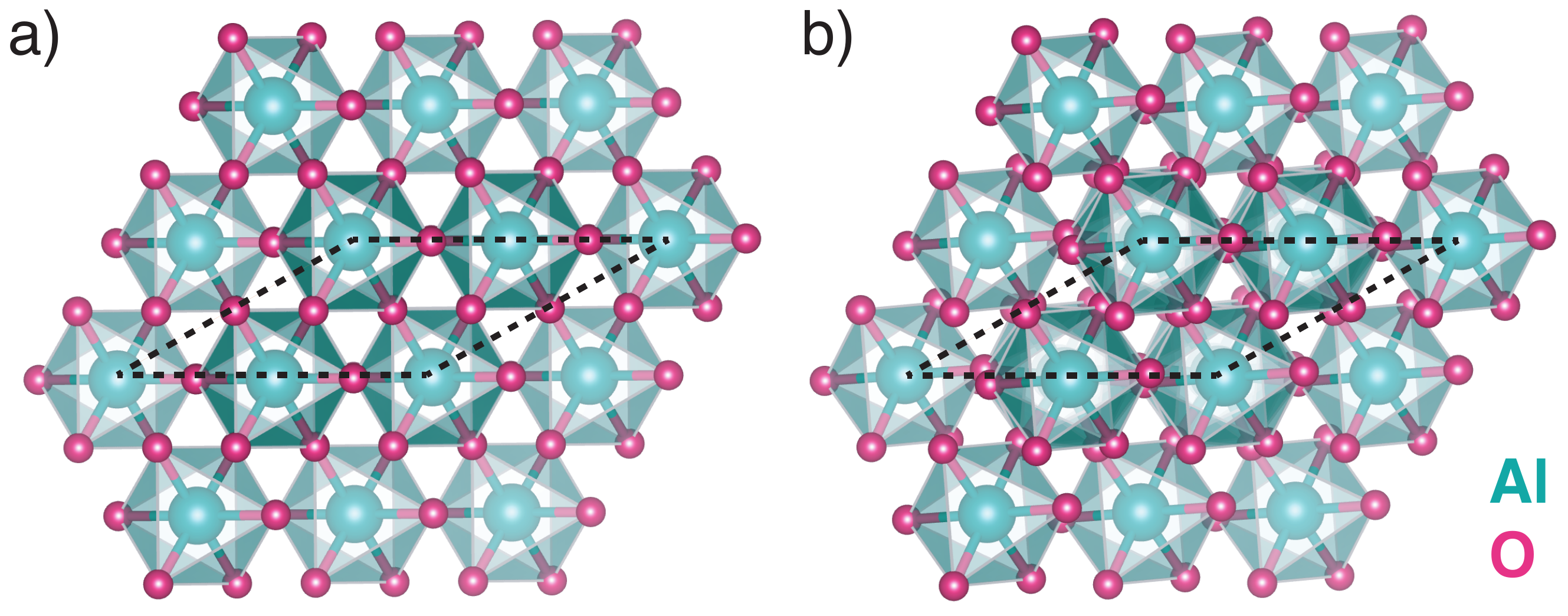}
    \caption{LaAlO$_3$ structures in the a) cubic $Pm\bar{3}m$ and b) rhombohedral $R\bar{3}c$ phases as viewed down the pseudocubic [111]$_{pc}$ axis. The lanthanum atoms have been omitted for clarity. Dashed lines indicate the unit cell. Adapted with permission from Ref. \onlinecite{Hayward2005Temperature}. Copyrighted by the American Physical Society.}
    \label{fig:structures}
\end{figure}

LaAlO$_3$ crystallizes in the rhombohedral $R\bar{3}c$ space group at room temperature and undergoes a structural phase transition to the cubic $Pm\bar{3}m$ space group at $\sim$813 K.\cite{Hayward2005Temperature} As shown in Figure \ref{fig:structures}, the AlO$_6$ octahedra in the rhombohedral phase are rotated by 5.1$^\circ$ in an anti-phase pattern ($a^-a^-a^-$ in Glazer notation\cite{Glazer1972Notation1, Glazer1975Notation2}) about the pseudocubic [111]$_{pc}$ axis. As the phase transition temperature is reached, the rotation of the AlO$_6$ octahedra is suppressed and the rotation angle drops to zero. The cubic phase can also be reached at room temperature by application of pressure; the rhombohedral phase transforms to $Pm\bar{3}m$ at 14.8 GPa.\cite{Guennou2011Pressure} Previous work has shown that both the temperature-induced\cite{lehnert00, Hayward2005Temperature,Scott1969frequencies} and pressure-induced\cite{Guennou2011Pressure} phase transitions are driven by a soft mode transforming like the irrep $R_4^+$ in the basis of the cubic $Pm\bar{3}m$ phase. This mode is associated with the rotation of the AlO$_6$ octahedra and becomes Raman-active with $\Gamma_1^+$ ($A_{1g}$) symmetry in the rhombohedral phase.

\begin{figure}
    \centering
    \includegraphics[width=8cm]{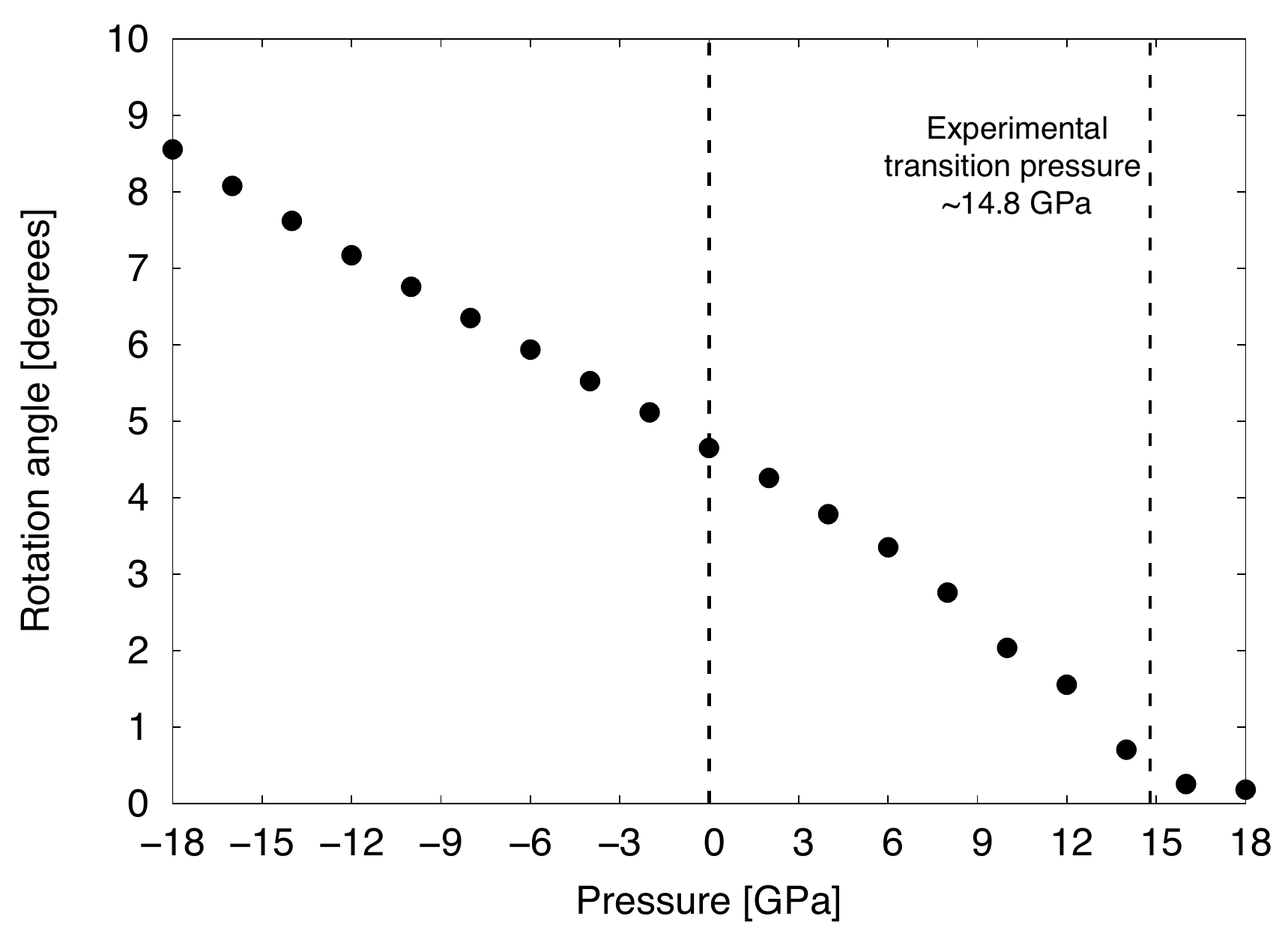}
    \caption{Variation in the angle of rotation of the AlO$_6$ octahedra with pressure in the $R\bar{3}c$ phase of LaAlO$_3$ from our first-principles calculations.}
    \label{fig:rotations}
\end{figure}

Figure \ref{fig:rotations} shows how the octahedral rotation angle varies with pressure in the rhombohedral phase from our first-principles calculations. Negative pressures obviously cannot be realized in experiments, however we include them here to gain a better understanding of trends across a wide range of pressures. Our results indicate that the rotation angle approaches zero close to the experimental transition pressure. Having established that we can qualitatively reproduce the structural parameters of LaAlO$_3$ under pressure, we now consider how $\alpha$ and the nonlinear phononics response can be tuned with pressure. 

\subsection{Vibrational Properties of Rhombohedral LaAlO$_3$}
Since the rhombohedral unit cell of LaAlO$_3$ contains 10 atoms there are 27 optical modes, which group theoretical analysis shows transform like the following irreps:
\begin{equation*}
    \Gamma_{\mathrm{optic}} = 2\Gamma_1^- \oplus 3\Gamma_2^- \oplus 5\Gamma_3^- \oplus \Gamma_1^+ \oplus 3\Gamma_2^+ \oplus 4\Gamma_3^+
\end{equation*}
Of these, the $\Gamma_2^-$ and $\Gamma_3^-$ modes are IR-active and the $\Gamma_1^+$ and $\Gamma_3^+$ modes are Raman-active; the remaining modes are silent. The $\Gamma_2^-$ modes are polarized along the same axis about which the AlO$_6$ octahedra rotate, whereas the $\Gamma_3^-$ modes are polarized perpendicular to this axis. Phonon frequencies from our first-principles calculations are in generally good agreement with experiment, as shown in Table \ref{tab_frequencies}. Group theoretical analysis further shows that the following mode couplings are allowed at cubic order in the lattice potential (quadratic in the IR-active mode and linear in the Raman mode): $\Gamma_2^-$ with $\Gamma_1^+$, $\Gamma_3^-$ with $\Gamma_1^+$ and $\Gamma_3^-$ with $\Gamma_3^+$. We considered the highest-symmetry direction for the doubly-degenerate $\Gamma_3^-$ and $\Gamma_3^+$ modes (this would lead to space group $Cc$ and $C2/c$ respectfully, if the modes were allowed to condense). In the following section, we investigate the strength of the nonlinear phononics response, that is, the average Raman displacement or the magnitude of $\alpha$, as a function of pressure for all phonon modes in LaAlO$_3$ corresponding to these three couplings.

\begin{center}
\begin{table}
\caption{Comparison between optical phonon frequencies in rhombohedral LaAlO$_3$ at zero pressure from our first-principles calculations, and from experiments.}
    \label{tab_frequencies}
    \centering
    \begin{tabular}{|c | c |  c |   }
        \hline
        Mode  &  DFT     & Experiment                   \\ 
        
         Symmetry  &  [cm$^{-1}$] & [cm$^{-1}$]         \\ 
        \hline
        $\Gamma_2^-$ ($A_{2u}$) &  &                    \\ 
         & 171 & 188 \cite{Willet-gies2014frequencies}  \\ 
         & 410 & 427 \cite{Willet-gies2014frequencies}  \\ 
         & 629 & 651 \cite{Willet-gies2014frequencies}  \\ 
         \hline
         $\Gamma_3^-$ ($E_u$)&  &                       \\ 
        
         & 185 & 188 \cite{Willet-gies2014frequencies}  \\ 
         & 297 & -   \\ 
         & 412 & 427 \cite{Willet-gies2014frequencies}  \\ 
         & 485 & 496 \cite{Willet-gies2014frequencies}  \\ 
         & 637 & 708 \cite{Willet-gies2014frequencies}  \\ 
        
        \hline
        $\Gamma_3^+$ ($E_g$)  &  &                      \\
         & 38  &  33 \cite{Scott1969frequencies}        \\ 
         & 150 & 152 \cite{Abrashev1999frequencies}     \\ 
         & 460 & 470 \cite{Abrashev1999frequencies}     \\ 
         & 463 & 487 \cite{Abrashev1999frequencies}     \\ 
        \hline
        $\Gamma_1^+$ ($A_{1g}$) & &                     \\
        & 134 & 123 \cite{Abrashev1999frequencies}      \\ 
        \hline
        $\Gamma_2^+$ ($A_{2g}$)  &  &                   \\ 
        & 142  & -                                      \\ 
        & 451  & -                                      \\
        & 747  & -                                      \\
        \hline
        $\Gamma_1^-$ ($A_{1u}$)  &  &                   \\
        & 325 &-                                        \\ 
        & 487  &  -                                     \\
        \hline
    \end{tabular}
\end{table}
\end{center}

\subsection{Pressure Dependence of Nonlinear Phononics Response}
\begin{figure}
    \centering
    \includegraphics[width=8cm]{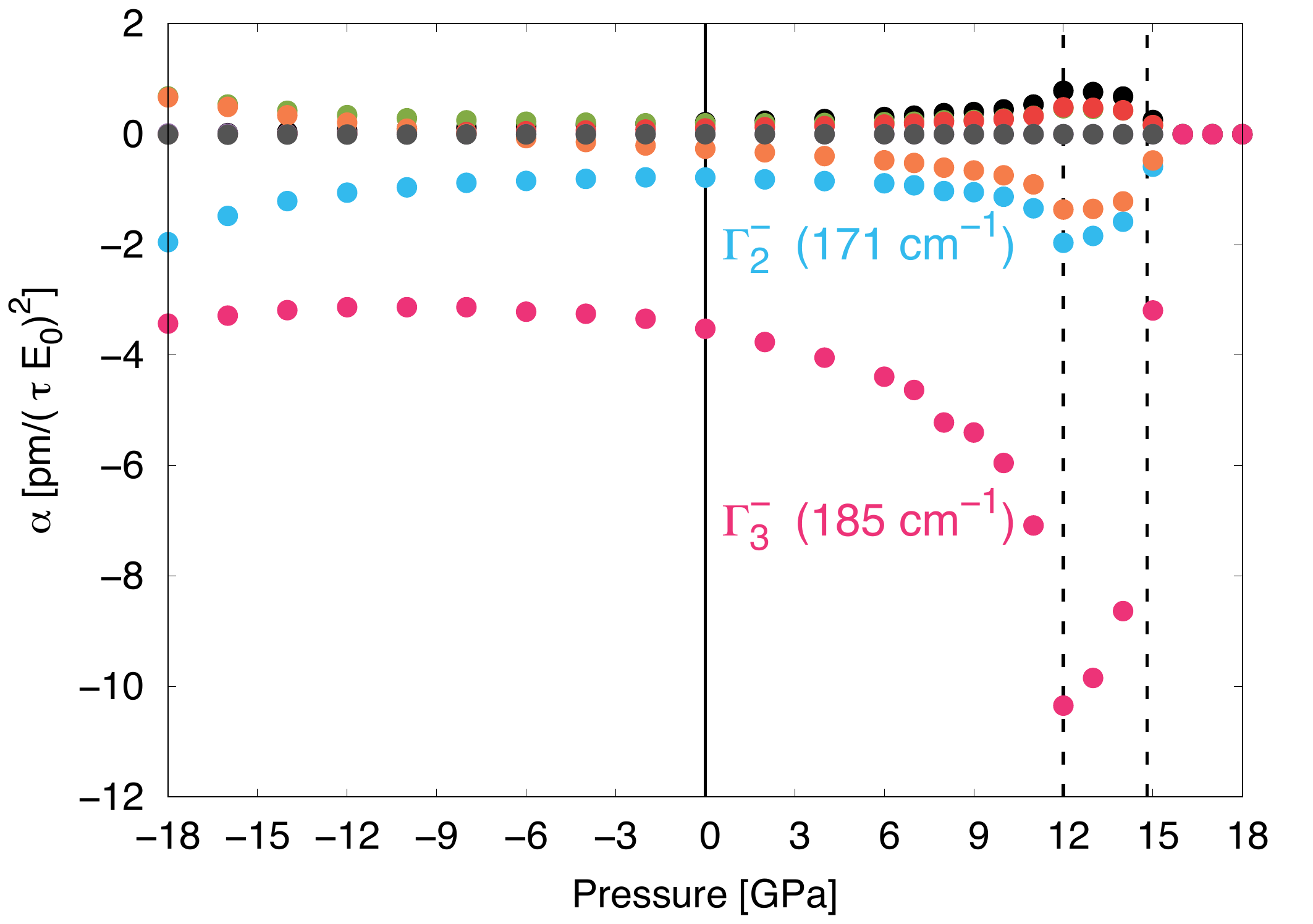}
    \caption{Variation in $\alpha$ (average Raman displacement) of the $\Gamma_1^+$ mode with pressure given excitation of one of the $\Gamma_2^-$ or $\Gamma_3^-$ modes. The vertical dashed line at 12 GPa indicates the pressure at which $\alpha$ is largest (for the $\Gamma_3^-$ mode at 185 cm$^{-1}$) and the vertical dashed line at 14.8 GPa indicates the experimental transition pressure for the rhombohedral -- cubic phase transition. These data appear in Table S1 of the Supplementary Information.}
    \label{gm1couplings}
\end{figure}

Figure \ref{gm1couplings} shows how $\alpha$, the average Raman displacement, varies as a function of pressure for coupling between the single $\Gamma_1^+$ mode and either the $\Gamma_2^-$ or $\Gamma_3^-$ IR-active modes. That is, Figure \ref{gm1couplings} illustrates how much the $\Gamma_1^+$ mode displaces given optical excitation of one of the IR-active modes in the material in a nonlinear phononics experiment. All of the quantities that appear in the figure of merit must be positive in a stable crystal except for $A$, and hence $A$ determines the sign of $\alpha$. The sign of $\alpha$ is somewhat arbitrary in the sense that it depends on which direction is considered a ``positive'' displacement of the Raman mode. We have two different criteria for selecting this direction. One, the positive Raman direction is chosen such that the coefficient of the third-order invariant of the Raman mode in the lattice potential is also positive. For example, given some potential $U_R = \frac{1}{2}K_RQ_R^2 + C_RQ_R^3$, we would define a positive $Q_R$ such that $C_R$ is positive. Two, in the case where there is no third-order invariant for the Raman mode, we choose the direction of the Raman mode such that a negative $\alpha$ indicates that the nonlinear phononics mechanism will unidirectionally displace the Raman mode towards a high-symmetry parent structure. Since the $\Gamma_1^+$ mode is already present in the equilibrium $R\bar{3}c$ phase (modes that transform like the identical representation are always allowed in space groups of any symmetry; these modes correspond to all motions of the atoms that are allowed by but do not change the space group symmetry) a negative sign of $\alpha$ means that the amplitude of the $\Gamma_1^+$ mode is reduced relative to its equilibrium amplitude in the rhombohedral phase, thus moving the system towards the cubic parent. 

We make note of a few key points. First, of the eight possible mode combinations, most result in an $\alpha$ that is insensitive to pressure and close to zero. This means that for excitation of most of the $\Gamma_2^-$ or $\Gamma_3^-$ modes, there is very little change in the $\Gamma_1^+$ amplitude and the response cannot really be tuned with pressure. Secondly, there are two mode combinations that result in significant displacement of the $\Gamma_1^+$ mode: excitation of a $\Gamma_2^-$ mode at 171 cm$^{-1}$ and excitation of a $\Gamma_3^-$ mode at 185 cm$^{-1}$. In both mode combinations, the average Raman displacement does vary with pressure and the magnitude of the displacement appears to be maximized close to the phase transition to the cubic phase. As an example of how to interpret the figure of merit we look at the $\Gamma_3^-/\Gamma_1^+$ combination for excitation of a $\Gamma_3^-$ mode at 185 cm$^{-1}$. At 0 GPa for a peak electric field of 1 MeV/cm with a one picosecond pulse-width, we predict an average peak displacement of the $\Gamma_1^+$ mode of -3.5 picometers, \textit{i.e.} the amplitude of the $\Gamma_1^+$ mode is reduced by 3.5 picometers compared to its equilibrium amplitude, as shown in Figure \ref{gm1couplings}. Going to 12 GPa, where there is an enhanced response (more negative $\alpha$), for the same peak electric field and pulse width we now expect a -10.3 picometer displacement of the $\Gamma_1^+$ mode, which is nearly triple that of the 0 GPa response.

\begin{figure}
    \centering
    \includegraphics[width=8cm]{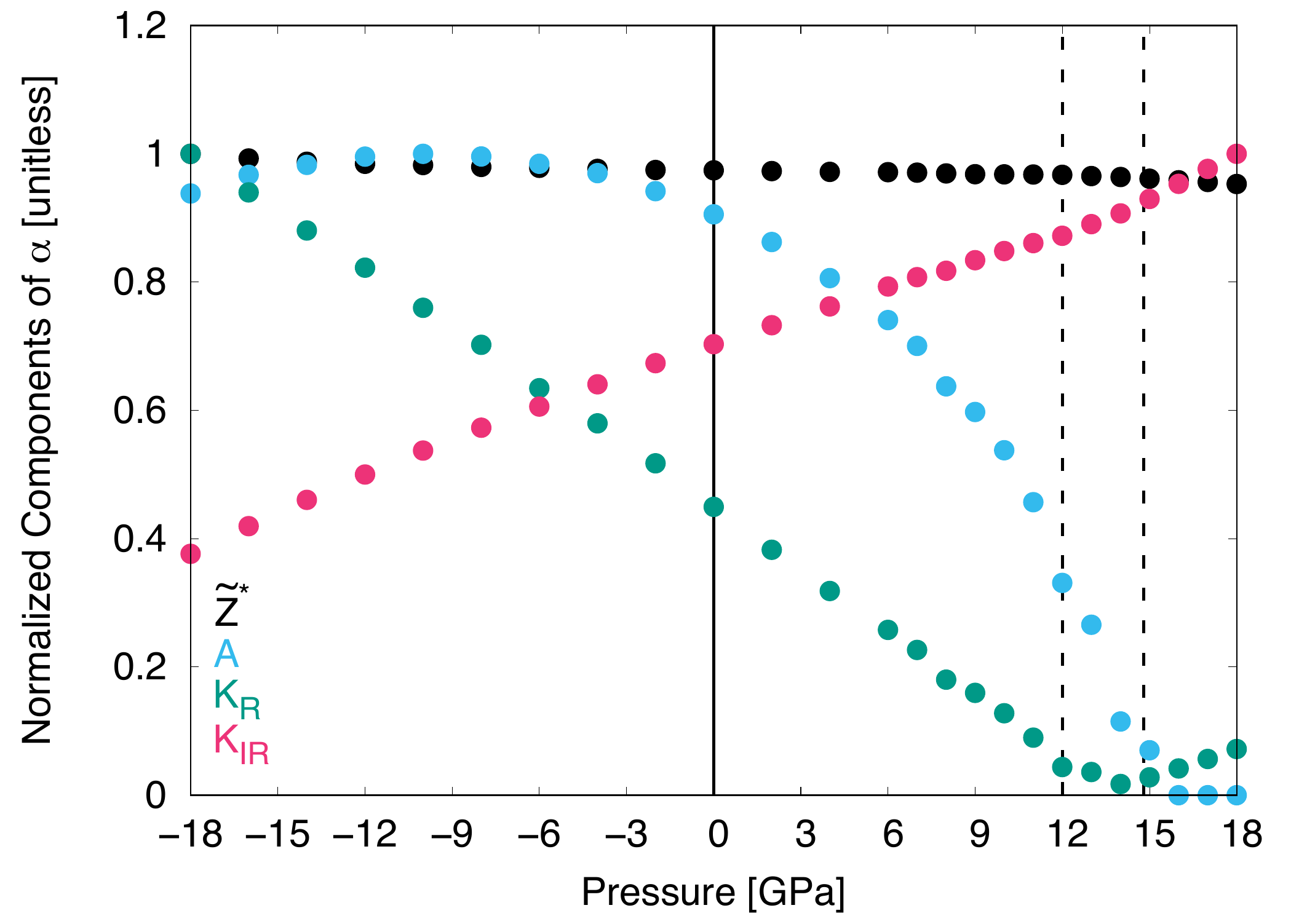}
    \caption{Variation in components of $\alpha$ with pressure for the $\Gamma_1^+$ mode given excitation of a $\Gamma_3^-$ mode at 185 cm$^{-1}$. Each component has been normalized by its peak value (we are plotting a ratio), so the maximum appears at 1. The vertical dashed line at 12 GPa indicates the pressure at which the magnitude of $\alpha$ is maximized (see Figure \ref{gm1couplings}), and the vertical dashed line at 14.8 GPa indicates the experimental transition pressure for the rhombohedral -- cubic phase transition. The dependence of $f_{IR}$ is shown in Figure S1 of the Supplementary Information.\cite{note3}}
    \label{gm1components}
\end{figure}

Equation \ref{alpha2} shows that a number of different intrinsic crystal properties contribute to $\alpha$ -- which of these makes the dominant contribution to the peak $\alpha$ for the $\Gamma_1^+$ mode? We focus on $\alpha$ for the $\Gamma_1^+$ mode given excitation of the $\Gamma_3^-$ phonon at 185 cm$^{-1}$, since Figure \ref{gm1couplings} shows that this combination results in the largest response. Figure \ref{gm1components} shows that $\widetilde{Z}^*$, the strength of the coupling between the IR mode and the light pulse, is essentially constant with pressure and therefore does not contribute to the large magnitude of the response at 12 GPa. In contrast, $A$ (the anharmonic coupling coefficient), the $\Gamma_1^+$ mode ($K_R$) and $\Gamma_3^-$ mode ($K_{IR}$) force constants all vary significantly with pressure. As the pressure increases towards the critical pressure (14.8 GPa) $K_R$ decreases contributing to an increase in the magnitude of $\alpha$. However, $K_{IR}$ increases and $A$ decreases over the same pressure range ($A$ is zero by symmetry in the cubic phase and so vanishes at the critical pressure). There is thus a competition between the growth in $\alpha$ due to $K_R$ decreasing with pressure, and a reduction in $\alpha$ due to decreasing $A$ and increasing $K_{IR}$; the details of this competition determine the pressure at which $\alpha$ is maximized and in this case $\alpha$ happens to be largest at 12 GPa.  Hence, in this case, it is not possible to attribute the peak response of $\alpha$ at 12 GPa to any one factor, rather it is the result of the balancing between these competing materials parameters. Interestingly, the large magnitude of $\alpha$ at 12 GPa is \emph{not} due to a particularly large $A$, which would indicate strong anharmonic coupling between the Raman-active and IR mode. It has generally been assumed that strong anharmonic coupling is required for a large nonlinear phononics response, but our results indicate that this is neither a necessary nor sufficient condition, since even if $A$ was large, a large $K_R$ and/or $K_{IR}$ could still make $\alpha$ small.

\begin{figure}
    \centering
    \includegraphics[width=8cm]{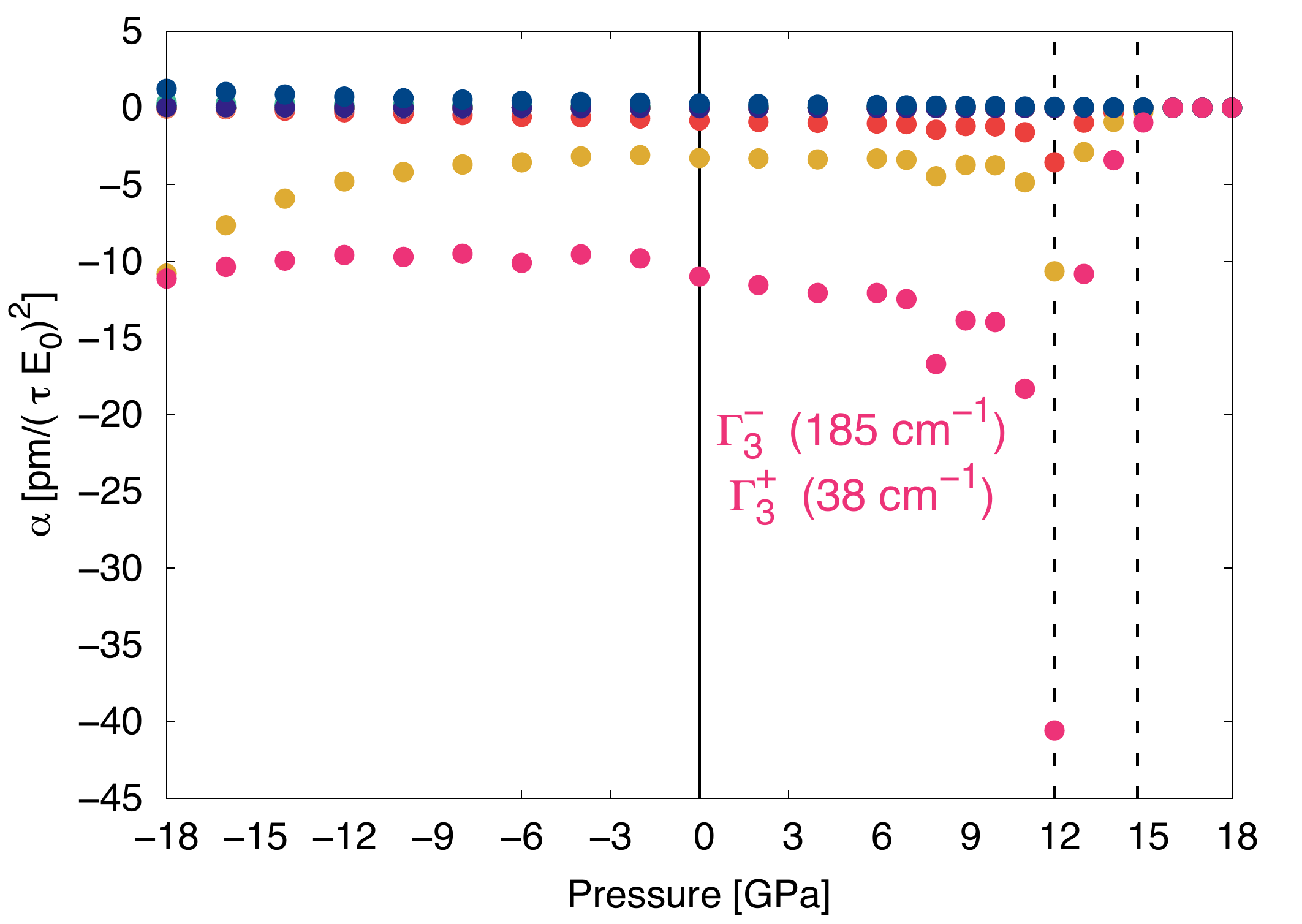}
    \caption{Variation in $\alpha$ (average Raman displacement) of the $\Gamma_3^+$ modes with pressure given excitation of one of the $\Gamma_3^-$ modes. The vertical dashed line at 12 GPa indicates the pressure at which $\alpha$ is largest (for the $\Gamma_3^-$ mode at 185 cm$^{-1}$) and the vertical dashed line at 14.8 GPa indicates the experimental transition pressure for the rhombohedral -- cubic phase transition. These data appear in Table S1 of the Supplementary Information.}
    \label{gm3couplings}
\end{figure}

We now turn our attention to the behavior of the $\Gamma_3^+$ Raman modes. Figure \ref{gm3couplings} shows how $\alpha$ varies as a function of pressure for coupling between the $\Gamma_3^+$ Raman-active modes and $\Gamma_3^-$ IR-active modes. Again, we note that of the twenty possible mode combinations, most are insensitive to pressure and result in an $\alpha$ that is close to zero. There is one mode combination that produces a significant displacement of the Raman-active mode: excitation of the same $\Gamma_3^-$ IR-active mode at 185 cm$^{-1}$ discussed above produces a large displacement of the Raman-active $\Gamma_3^+$ mode at 38 cm$^{-1}$. In particular, note the difference in scale of the $y$-axis between Figures \ref{gm1couplings} and \ref{gm3couplings}. Can this large displacement be attributed to any one component of $\alpha$?

\begin{figure}
    \centering
    \includegraphics[width=8cm]{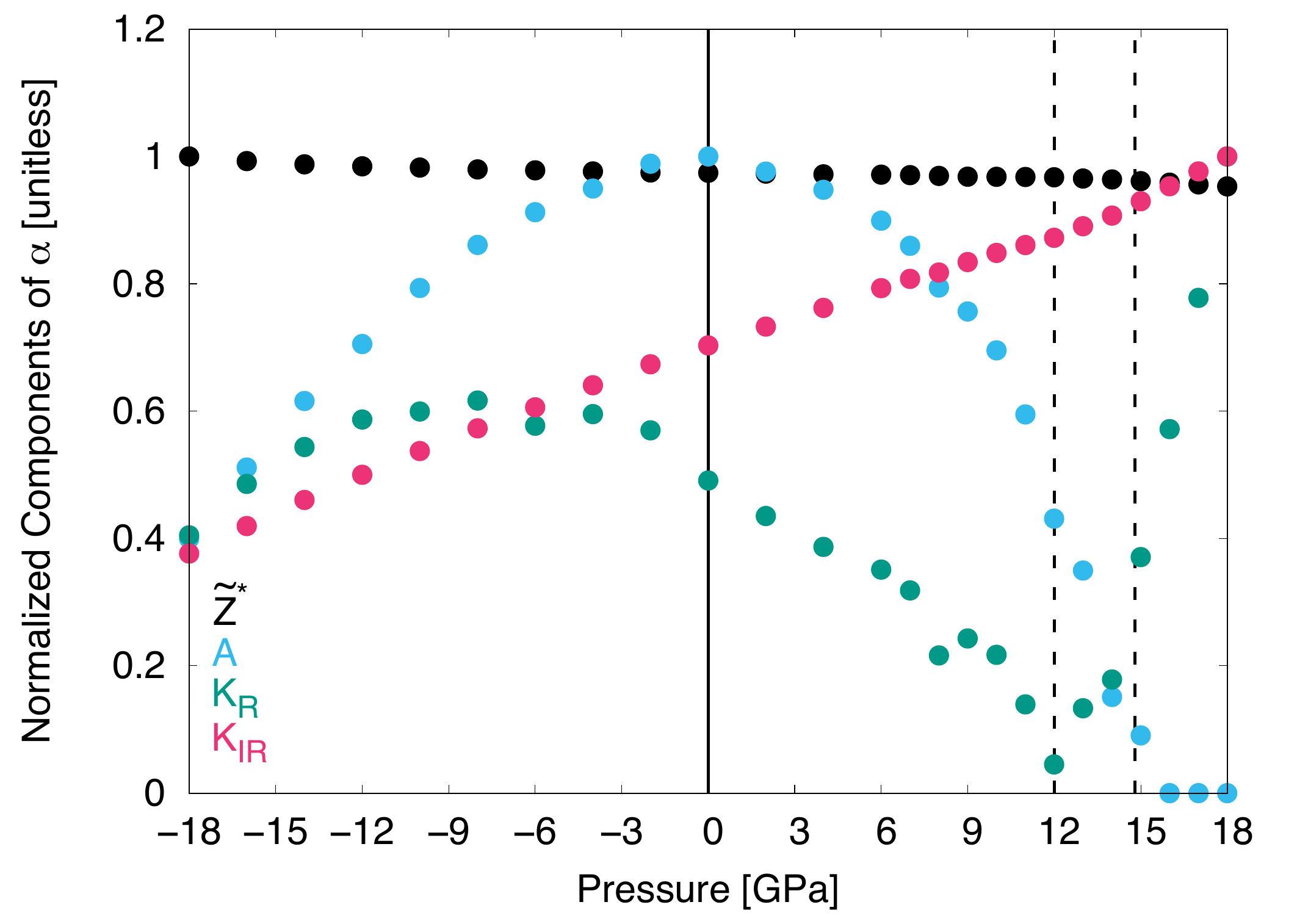}
    \caption{Variation in components of $\alpha$ with pressure for the 38 cm$^{-1}$ $\Gamma_3^+$ mode given excitation of a $\Gamma_3^-$ mode at 185 cm$^{-1}$. Each component has been normalized by its peak value (we are plotting a ratio), so the maximum appears at 1. The vertical dashed line at 12 GPa indicates the (positive) pressure at which the magnitude of $\alpha$ is maximized (see Figure \ref{gm3couplings}), and the vertical dashed line at 14.8 GPa indicates the experimental transition pressure for the rhombohedral -- cubic phase transition. The dependence of $f_{IR}$ is shown in Figure S2 of the Supplementary Information.\cite{note3}}
    \label{gm3components}
\end{figure}

Since the $\Gamma_3^-$ mode involved in this coupling is the same as discussed above with the $\Gamma_1^+$ coupling, the behavior of $\widetilde{Z}^*$ and $K_{IR}$ with pressure is the same, as a comparison of Figures \ref{gm1components} and \ref{gm3components} shows. The value of $\alpha$ is actually fairly constant with increasing pressure until close to 12 GPa -- an increase in $K_{IR}$ with pressure and a decrease in $A$, both of which would make the magnitude of $\alpha$ smaller, are countered by a decrease in $K_R$. There is a sharp drop in $\alpha$ close to 12 GPa however and this is due to a very small $K_R$. At zero pressure, $K_R$ for the $\Gamma_3^+$ mode is 0.08 eV/\AA$^2$, it then decreases with pressure and reaches a minimum value of 0.008 eV/\AA$^2$ at 12 GPa. Hence, in the case of the $\Gamma_3^-/\Gamma_3^+$ mode combination, $\alpha$ is maximized at 12 GPa due to a very small $K_R$. Note that although the force constant for the $\Gamma_1^+$ mode decreases proportionally over the same pressure range, it is a much harder mode to begin with, 1.06 eV/\AA$^2$ at zero pressure, and reaches a value of 0.10 eV/\AA$^2$ at 12 GPa, an order of magnitude larger than $K_R$ for the $\Gamma_3^+$ mode at the same pressure. As an aside, we also emphasize again that the largest $A$ does not necessarily correspond to the largest Raman response --  $A$ for the $\Gamma_3^-/\Gamma_3^+$ combination is nearly four times smaller (-0.52 eV/\AA$^3$) than $A$ for the $\Gamma_3^-/\Gamma_1^+$ combination (-2.12 eV/\AA$^3$) discussed above.

Our results for the $\Gamma_3^-/\Gamma_1^+$ and $\Gamma_3^-/\Gamma_3^+$ mode couplings indicate that the magnitude of the structural change induced by the nonlinear phononics mechanism is not just material dependent but also mode dependent. That is, depending on the IR-active mode that is optically excited in a given material it is possible to have a large response (large structural change) or also very little response. In addition, the intrinsic materials factors that underlie a large response can also differ for different mode couplings in the \emph{same} material. Further work in this area should help experimentalists identify the most promising materials for investigation, and within those materials, the most promising mode combinations.


\section{Summary and Conclusions}
Our work demonstrates that extrinsic variables such as pressure (and likely, epitaxial strain) can effectively tune the magnitude of structural change induced by the nonlinear phononics mechanism in an ultrafast optical experiment. For LaAlO$_3$ in particular we predict a large (compared to ambient pressure) structural change around 12 GPa due to displacements of the $\Gamma_1^+$ and $\Gamma_3^+$ Raman-active modes following excitation of a $\Gamma_3^-$ IR-active mode. Based on recent high pressure ultrafast studies,\cite{Braun2018PressureV02} we believe pressure will be a useful tool for the study and enhancement of the nonlinear phononics effect in materials like LaAlO$_3$.  We elucidated the origin of these responses by defining a figure of merit based on intrinsic materials properties. In the case of the $\Gamma_3^-/\Gamma_1^+$ mode combination, the enhanced response at 12 GPa is due to a combination of a soft $\Gamma_1^+$ force constant and a sizeable anharmonic coupling coefficient $A$. In the case of the $\Gamma_3^-/\Gamma_3^+$ mode combination, the enhanced response at 12 GPa is due to a very small $\Gamma_3^+$ force constant. 

We speculate that, as is usually the case in condensed matter systems, the magnitude of structure switching achievable via nonlinear phononics will be amplified close to the boundary of a structural phase transition. Many complex oxides, and perovskites in particular, undergo structural phase transitions, making them a promising class of materials for continued exploration. Our figure of merit applies generally to any crystal with the lattice potential shown above and most of the terms are experimentally measurable with IR and Raman spectroscopy. The anharmonic coupling coefficient is not easily obtained from experiments, however it can be readily calculated from first-principles. In fact, since all of the intrinsic quantities in the figure of merit can be calculated from first-principles, high-throughput computational studies are feasible and may be an efficient approach to identifying materials that exhibit large structural changes in nonlinear phononics experiments.

\acknowledgements
Initial work on this project was performed by G. K. and supported by the National Science Foundation under award DMR-1719875. J. Z. K. and N. A. B. were supported by the Department of Energy -- Office of Basic Energy Sciences under award DE-SC0019414. This research used resources of the National Energy Research Scientific Computing Center (NERSC), a U.S. Department of Energy Office of Science User Facility located at Lawrence Berkeley National Laboratory, operated under Contract No. DE-AC02-05CH11231.

\bibliography{citations.bib}

\begin{thebibliography}{10}
\bibitem{note2}
Short pulse widths will contain a broad band of frequencies and effects such as
  impulsive Raman scattering may begin to dominate.
\bibitem{note}
This approximation will break down if the coupling between the IR and Raman
  modes is very strong.
\end{thebibliography}
\bibliographystyle{ieeetr}

\begin{widetext}
\clearpage
\end{widetext}

\newcommand{\beginsupplement}{%
        \setcounter{table}{0}
        \renewcommand{\thetable}{S\arabic{table}}%
        \setcounter{figure}{0}
        \renewcommand{\thefigure}{S\arabic{figure}}%
     }

\beginsupplement

\onecolumngrid
\begin{center}
\textbf{A Strategy to Identify Materials Exhibiting a Large Nonlinear Phononics Response:\\ Tuning the Ultrafast Structural Response of LaAlO$_3$ with Pressure}

Jeffrey Z. Kaaret,$^1$ Guru Khalsa$^2$ and Nicole A. Benedek$^2$

1. School of Applied and Engineering Physics and 2. Department of Materials Science and Engineering, Cornell University, Ithaca NY 14853, USA
\end{center}
\twocolumngrid

\begin{figure}[h]
    \centering
    \includegraphics[width=8cm]{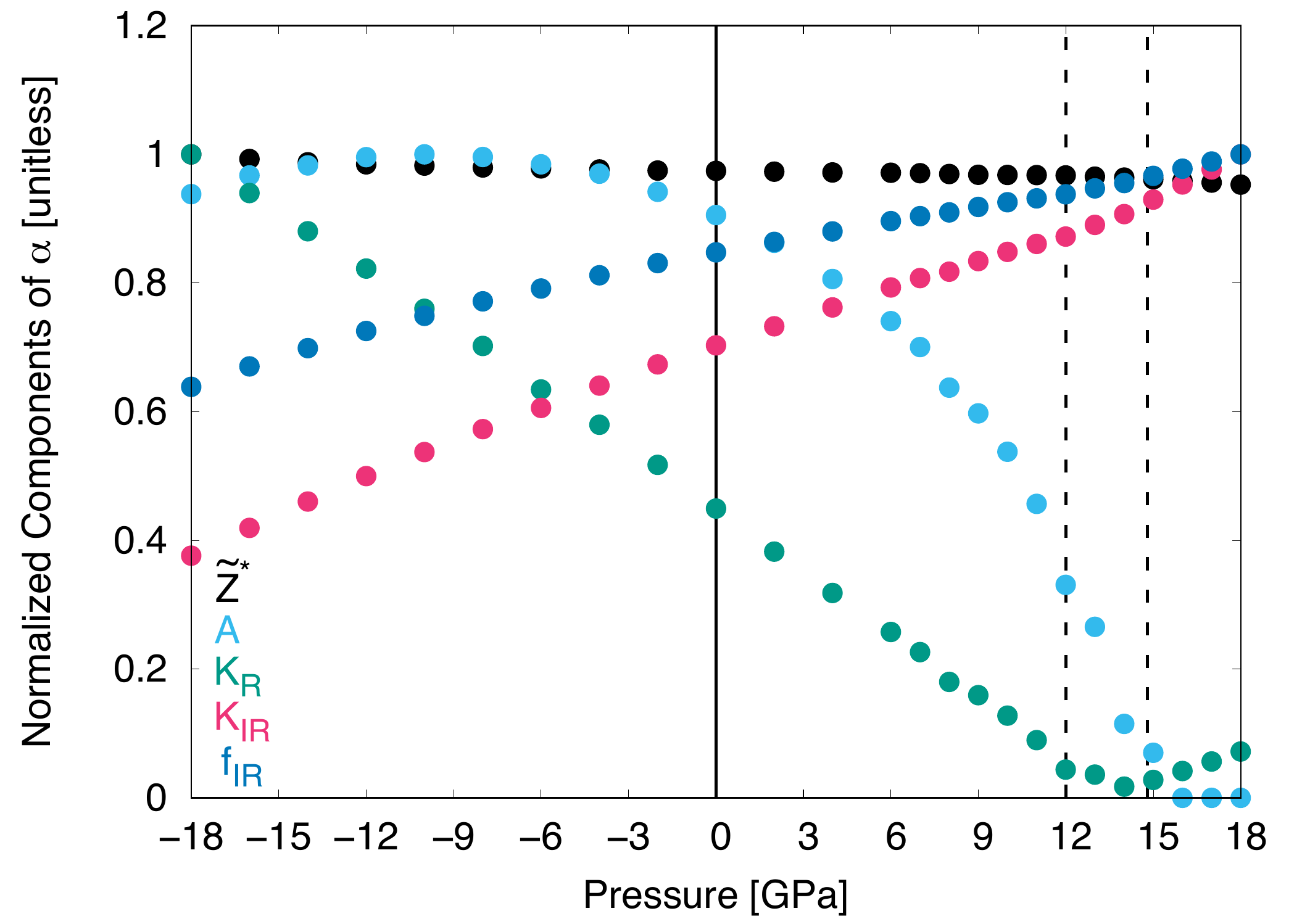}
    \caption{Figure 5 with the $f_{IR}$ dependence. Variation in components of $\alpha$ with pressure for the $\Gamma_1^+$ mode given excitation of a $\Gamma_3^-$ mode at 185 cm$^{-1}$. Each component has been normalized by its peak value (we are plotting a ratio), so the maximum appears at 1. The vertical dashed line at 12 GPa indicates the pressure at which the magnitude of $\alpha$ is maximized, and the vertical dashed line at 14.8 GPa indicates the experimental transition pressure for the rhombohedral -- cubic phase transition. } 

    \label{fig:Components_w_freq_GM1+}
\end{figure}

\newpage

\begin{figure}[h]
    \centering
    \includegraphics[width=8cm]{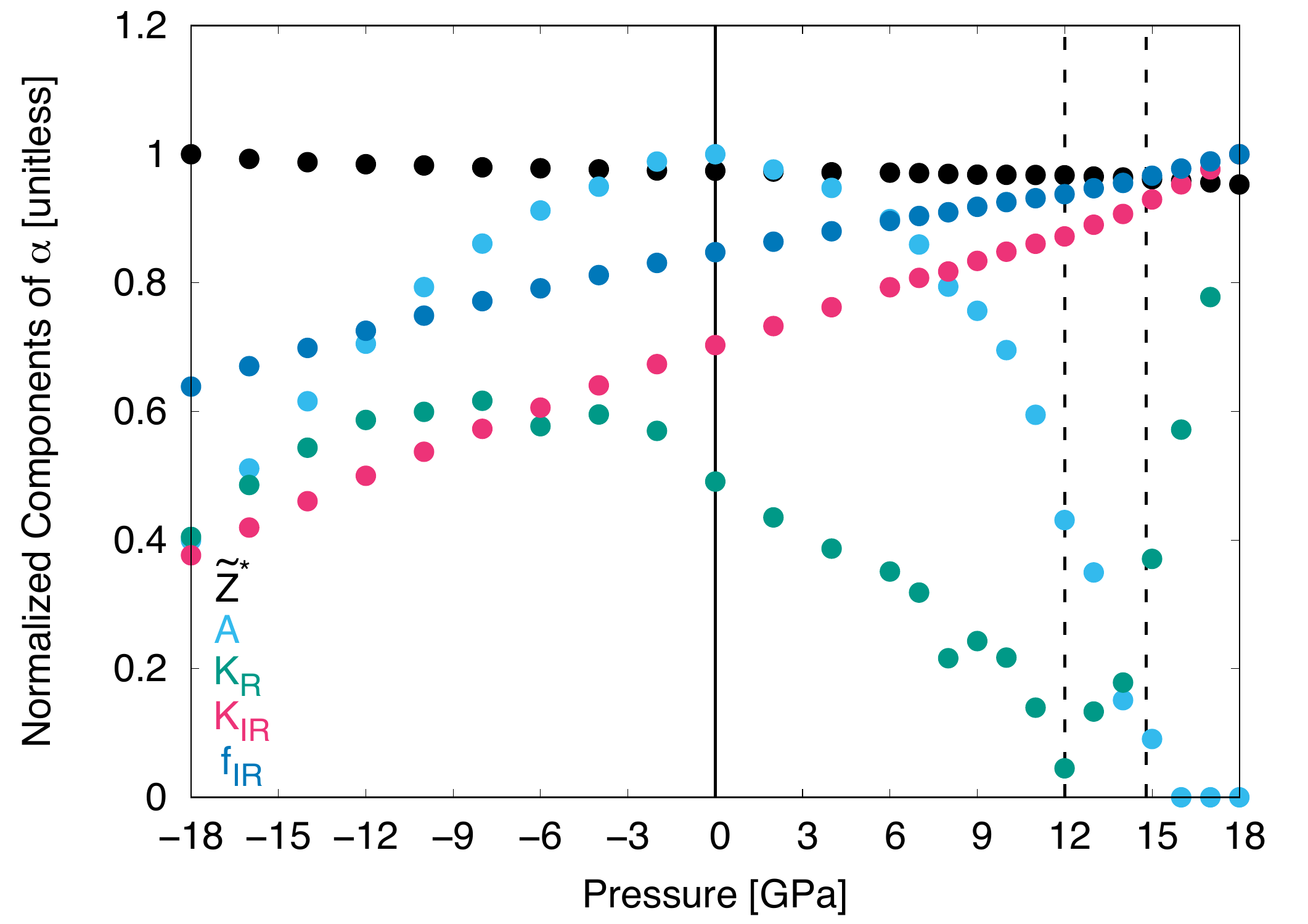}
    \caption{Figure 7 with the $f_{IR}$ dependence. Variation in components of $\alpha$ with pressure for the 38 cm$^{-1}$ $\Gamma_3^+$ mode given excitation of a $\Gamma_3^-$ mode at 185 cm$^{-1}$. Each component has been normalized by its peak value (we are plotting a ratio), so the maximum appears at 1. The vertical dashed line at 12 GPa indicates the (positive) pressure at which the magnitude of $\alpha$ is maximized, and the vertical dashed line at 14.8 GPa indicates the experimental transition pressure for the rhombohedral -- cubic phase transition. }

    \label{fig:Components_w_freq_GM3+}
\end{figure}

\begin{center}
 \begin{table*}[hbt!]
\caption{Relevant quantities (Raman and IR force constants, mode effective charge, anharmonic coupling, and figure of merit) for all couplings studied in this paper at 0 applied pressure. The final two columns are the positive peak pressure for each mode coupling and the corresponding peak figure of merit. This is to give scale of the figure of merit for the couplings we did not discuss in depth. $E_0$ is in units of MV/cm, $\tau$ is in units of picoseconds. *These $\Gamma_3^+$ modes cross in frequency resulting in a hybridization near 0 GPa, as a result we have interpolated the anharmonic coupling terms related to these modes as they are noisy about 0 GPa. }
    \label{tab_with_freq_quantities}   
    
    \begin{tabular}{|c | c | c |  c | c | r | r | r | r | r | r |  }
        \hline
        Raman & Raman Frequency & K$_R$  & IR & IR Frequency  & K$_{IR}$   & MEC     &  A  &  $\alpha$ & Peak Pressure & Peak $\alpha$ \\
        IRREP & [cm$^{-1}$]  & [eV/\AA$^2$] & IRREP & [cm$^{-1}$] & [eV/\AA$^2$] & [e$^-$] & [eV/\AA$^3$] &  [pm/($\tau E_0$)$^2$] & [GPa] &  [pm/($\tau E_0$)$^2$] \\  
        \hline
        $\Gamma_1^+$ & 134 &  1.06 & $\Gamma_2^-$ & 629 & 24.25 & 2.48 &  3.09 &  2.2$\times$10$^{-1}$ & 12 &  7.9$\times$10$^{-1}$ \\   	
        & &                        &              & 410 & 13.50 & 5.83 &  0.37 &  2.0$\times$10$^{-1}$ & 12 &  4.7$\times$10$^{-1}$ \\    
        & &                        &              & 171 &  2.68 & 5.41 & -0.39 & -7.9$\times$10$^{-1}$ & 12 & -2.0$\times$10$^{+0}$ \\ \cline{4-11}  
        & &                        & $\Gamma_3^-$ & 637 & 24.82 & 2.37 &  1.66 &  1.0$\times$10$^{-1}$ & 12 &  4.9$\times$10$^{-1}$ \\    
        & &                        &              & 485 & 23.32 & 0.37 &  1.51 &  1.5$\times$10$^{-3}$ &  0 &  1.5$\times$10$^{-3}$ \\ 
        & &                        &              & 412 & 13.55 & 6.06 & -0.46 & -2.6$\times$10$^{-1}$ & 12 & -1.4$\times$10$^{+0}$ \\ 
        & &                        &              & 297 &  5.20 & 0.17 & -1.00 & -1.6$\times$10$^{-3}$ &  0 & -1.6$\times$10$^{-3}$ \\ 
        & &                        &              & 185 &  3.16 & 5.35 & -2.12 & -3.5$\times$10$^{+0}$ & 12 & -1.0$\times$10$^{+1}$ \\ 
        \hline
       $\Gamma_3^+$* & 463 & 12.42 & $\Gamma_3^-$ & 637 & 24.82 & 2.37 &  0.79 &  4.1$\times$10$^{-3}$ &  0 &  4.1$\times$10$^{-3}$ \\         
        & &                        &              & 485 & 23.32 & 0.37 & -0.88 & -7.3$\times$10$^{-5}$ &  0 & -7.3$\times$10$^{-5}$ \\         
        & &                        &              & 412 & 13.55 & 6.06 &  0.89 &  4.3$\times$10$^{-2}$ &  0 &  4.3$\times$10$^{-2}$ \\         
        & &                        &              & 297 &  5.20 & 0.17 &  0.45 &  6.1$\times$10$^{-5}$ &  0 &  6.1$\times$10$^{-5}$ \\         
        & &                        &              & 185 &  3.16 & 5.35 &  0.96 &  1.3$\times$10$^{-1}$ &  0 &  1.3$\times$10$^{-1}$ \\         
        \hline        
       $\Gamma_3^+$* & 460 & 14.61 & $\Gamma_3^-$ & 637 & 24.82 & 2.37 & -1.18 & -6.3$\times$10$^{-3}$ &  0 & -6.3$\times$10$^{-3}$ \\         
        & &                        &              & 485 & 23.32 & 0.37 &  2.43 &  2.0$\times$10$^{-4}$ &  0 &  2.0$\times$10$^{-4}$ \\         
        & &                        &              & 412 & 13.55 & 6.06 & -0.66 & -3.2$\times$10$^{-2}$ &  0 & -3.2$\times$10$^{-2}$ \\         
        & &                        &              & 297 &  5.20 & 0.17 & -0.45 & -6.2$\times$10$^{-5}$ &  0 & -6.2$\times$10$^{-5}$ \\         
        & &                        &              & 185 &  3.16 & 5.35 & -0.06 & -8.6$\times$10$^{-3}$ &  2 & -9.4$\times$10$^{-3}$ \\         
        \hline   
        $\Gamma_3^+$ & 150 & 10.45 & $\Gamma_3^-$ & 637 & 24.82 & 2.37 &  0.58 &  3.7$\times$10$^{-3}$ &  0 &  3.7$\times$10$^{-3}$ \\         
        & &                        &              & 485 & 23.32 & 0.37 &  0.12 &  1.2$\times$10$^{-5}$ &  0 &  1.2$\times$10$^{-5}$ \\         
        & &                        &              & 412 & 13.55 & 6.06 &  0.47 &  2.7$\times$10$^{-2}$ &  0 &  2.7$\times$10$^{-2}$ \\         
        & &                        &              & 297 &  5.20 & 0.17 &  0.73 &  1.2$\times$10$^{-4}$ &  0 &  1.2$\times$10$^{-4}$ \\         
        & &                        &              & 185 &  3.16 & 5.35 &  1.70 &  2.9$\times$10$^{-1}$ &  0 &  2.9$\times$10$^{-1}$ \\         
        \hline   
        $\Gamma_3^+$ &  38 &  0.08 & $\Gamma_3^-$ & 637 & 24.82 & 2.37 & -1.04 & -8.3$\times$10$^{-1}$ & 12 & -3.6$\times$10$^{+0}$ \\         
        & &                        &              & 485 & 23.32 & 0.37 &  0.45 &  5.7$\times$10$^{-3}$ &  0 &  5.7$\times$10$^{-3}$ \\         
        & &                        &              & 412 & 13.55 & 6.06 & -0.45 & -3.3$\times$10$^{+0}$ & 12 & -1.1$\times$10$^{+1}$ \\         
        & &                        &              & 297 &  5.20 & 0.17 & -0.41 & -8.4$\times$10$^{-3}$ &  0 & -8.4$\times$10$^{-3}$ \\         
        & &                        &              & 185 &  3.16 & 5.35 & -0.52 & -1.1$\times$10$^{+1}$ & 12 & -4.1$\times$10$^{+1}$ \\          
        \hline   
    \end{tabular}
\end{table*}    
\end{center}

\end{document}